\documentclass[%
 reprint,
superscriptaddress,
 aps,pre,
showkeys 
]{revtex4-2}

\usepackage[utf8]{inputenc}
\usepackage[english]{babel}
\usepackage{color}
\usepackage{graphicx}%
\usepackage{bm}%
\usepackage[export]{adjustbox}%
\usepackage{float}
\usepackage{lipsum}
\usepackage{mwe}
\usepackage{amsmath}
\usepackage{amssymb}
\usepackage{framed}
\usepackage{xcolor}
\colorlet{shadecolor}{gray!18}

\def\be{\begin{equation}}
\def\eea{\end{eqnarray}}
\def\ee{\end{equation}}
\def\bea{\begin{eqnarray}}
\def\ea{\end{array}}
\def\ba{\begin{array}}

\newcommand{\sign}{\operatorname{sign}}

\begin{document}

\title{
  Lévy walkers inside spherical shells with absorbing boundaries:\\
  Towards settling the optimal Lévy walk strategy  for random searches}

\author{L. G. P. Caramês}
\affiliation{Department of Physics, Federal University of 
Rio Grande do Norte, 59078-900 Natal-RN, Brazil}

\author{Y. B. Matos}
\affiliation{Departamento de Física, Universidade Federal do Paraná,
  81531-980 Curitiba-PR, Brazil}

\author{F. Bartumeus}
\affiliation{Centre d'Estudis
  Avan\c{c}ats de Blanes-CEAB-CSIC, Girona 17300, Spain}
  
\affiliation{CREAF,
  Universitat Aut\`onoma de Barcelona, Cerdanyola del Vall\`es 08193,
  Spain}
  
\affiliation{ICREA, Institució Catalana de Recerca
  i Estudis Avançats, Barcelona 08010, Spain}
 
\author{C. G. Bezerra}
\affiliation{Department of Physics, Federal University of 
Rio Grande do Norte, 59078-900 Natal-RN, Brazil}

\author{T.~Macrì}    
\affiliation{Department of Physics, Federal University of 
Rio Grande do Norte, 59078-900 Natal-RN, Brazil}

\affiliation{ITAMP, Harvard-Smithsonian Center for Astrophysics,
  Cambridge, Massachusetts 02138, USA}

\author{M. G. E. da Luz}
\affiliation{Departamento de Física, Universidade Federal do Paraná,
  81531-980 Curitiba-PR, Brazil}    

    \author{E. P. Raposo}
\affiliation{Laboratório de Física
  Te\'orica e Computacional, Departamento de Física, Universidade
  Federal de Pernambuco, Recife-PE 50670-901, Brazil}

\author{G. M. Viswanathan}
\affiliation{Department of Physics, Federal University of 
Rio Grande do Norte, 59078-900 Natal-RN, Brazil}

\affiliation{National Institute of Science and
  Technology of Complex Systems,
  Federal University of 
Rio Grande do Norte, 59078-900 Natal, RN, Brazil}

\begin{abstract}

The Lévy flight foraging hypothesis states that organisms must have
evolved adaptations to exploit Lévy walk search strategies.
Indeed, it is widely accepted that inverse square Lévy walks
optimize the search efficiency in foraging with unrestricted
revisits (also known as non-destructive foraging).
However, a mathematically rigorous demonstration of this for
dimensions $D \geq 2$ is still lacking.
Here we study the very closely related problem of a Lévy walker
inside annuli or spherical shells with absorbing boundaries.
In the limit that corresponds to the foraging with unrestricted
revisits, we show that inverse square Lévy walks optimize the
search.
This constitutes the strongest formal result to date supporting
the optimality of inverse square Lévy walks search strategies.

\end{abstract}

\maketitle

Lévy stable distributions, Lévy walks and flights, have attracted
wide attention since the 1990s in areas as diverse as particle
kinetics~\cite{ea,eb} and random lasers~\cite{ec,ed}.
In particular, almost a quarter of a century ago it was
proposed~\cite{nature1999} that Lévy walks with an inverse
square law distribution for the step lengths can lead to optimal
search strategies since they maximize the encounter rate with
sparse, randomly distributed,
revisitable targets when the search restarts in the vicinity
of the previously visited target (thus available for further
visits) and with no information about the past behavior ---
an uninformed process.
This key fact about ``non-destructive foraging'' has
motivated the formulation of the Lévy flight foraging hypothesis
(LFH) in ecology which holds that for many species such 
optimization may have led to adaptations for Lévy walk search
strategies~\cite{phys-life-rev2015,plos-comput-biol2017}.  
When originally published, the result caused surprise, because
it questioned and then overturned the assumption that organisms
move solely according to Brownian motion.
In the last couple of decades, however, these results have
been exaustively verified in many different
instances~\cite{pnas2008},
becoming widely accepted (see, e.g., Ref. \cite{foraging-book}).

More recently there has been renewed interest in certain
fundamental and formal aspects of the problem.
For example, eventual theoretical findings against the
optimality of inverse square Lévy walk searches for any spatial
dimensions $D$ \cite{ref1} have been shown not to be
applicable to the paradigmatic non-destructive random
search context \cite{comment}.
But this debate is understandable given that although several
concrete situations have pointed to the aforementioned
optimality, it is mathematically very hard to prove (or
disprove) and no general developments have appeared so far
in the literature.

Aiming to provide an important advance towards finally settling
positively the issue, here we study a very closely related problem
that is helpfully much easier to deal with.
Specifically, we investigate in detail a Lévy walker inside a
2-dimensional annulus with absorbing boundaries.
The analysis is also valid for higher dimensions by considering
hyperspherical shells.
The essential point is that the original foraging problem
can thus be analyzed indirectly by proxy, through characterizing
the first passage time (FPT) in annuli and shell geometries.
Indeed, the first passage time can be thought as the basic
building block, representing the finding of successive targets
in the random search for many targets, see
Fig.~\ref{fig:annuli-schema}.
The inner boundary represents the previously found target
site in the foraging problem.
The outer boundary replaces all other targets.

As shown below, our main result is that inverse square Lévy
walks strategies become optimal for extremizing the mean FPT
when the initial position becomes arbitrarily close to the
inner radius.
In other words, the first passage time for a Lévy walker to reach
the boundary of the annulus or shell is minimized under certain
initial conditions that correspond to the case of foraging
with unrestricted revisits, i.e., for the non-destructive
limit of the original foraging problem.

{But before continuing, a few words about commonly used jargon
would be in order.
The traditional definition of  ``non-destructive foraging'' seems
not to capture the most general conditions for the relevant
search scenario here.
Instead, we adopt the term ``foraging with (unrestricted) revisits''.
Here by
``unrestricted''  is meant, e.g., that there is no regenerative (or
waiting) times \cite{raposo-2003,santos-2004} to approach any
previously visited target.
Nor some targets are more difficult to reach than others,
apart from their Euclidean distance to the walker.
Although a subtle difference, we shall make sure that our
analysis encompasses a large number of instances, so that
the revisits could occur for several reasons besides just
retracking regions or areas previously scanned:
sensory errors, re-emergent or replenishable targets,
heterogeneous target distributions in patches that can be
exploited in several visits, and so forth.
}

The foraging model and the model of search inside annuli and spherical
shells are presented in Section \ref{sec-model}. Sections
\ref{sec-ana} and \ref{sec-num} present analytical and numerical
results, respectively. We end with concluding remarks in Section \ref{sec-con}.

  \section{Models}  \label{sec-model}
\subsection{\em Foraging model:} 
The foraging model consists of a general strategy rule
for the walker (of unit velocity, so that the distance
traveled $L=t$, with $t$ the traveled time) to search
for targets randomly distributed in a $D$-dimensional
space (Fig. \ref{fig:annuli-schema} (a) and (b)):\\
(i) If there are targets within
a radius of detection (or 
sight range) $r_{v}$,
then the searcher moves in a straight line to the nearest
target.
\\
(ii) If there is no target at distance $r_{v}$ from the
searcher, then for its $j$-th step, the searcher chooses
a random direction and draws a distance from a power law
tailed probability distribution
\begin{equation}
  p\left(\ell_{j}\right) \sim 1/\ell_{j}^{\mu},
  \label{levy_dist}
\end{equation}
So, the walker starts moving incrementally, continually
looking for a new target within sight radius $r_{v}$
along the way.
If no target is detected, the searcher stops after
covering the distance $\ell_{j}$ and (ii) resumes.
Otherwise, it proceeds according to rule (i).

The time-averaged search efficiency $\eta$ is defined as 
\begin{equation}
  \eta=\lim_{t\rightarrow\infty}N/t,
  \label{eff_ratio}
\end{equation}
where $N$ is the number of targets found  in time $t$.

\subsection{Search inside annuli and spherical shells}

\begin{figure}[t]
    \includegraphics[width=1\linewidth]{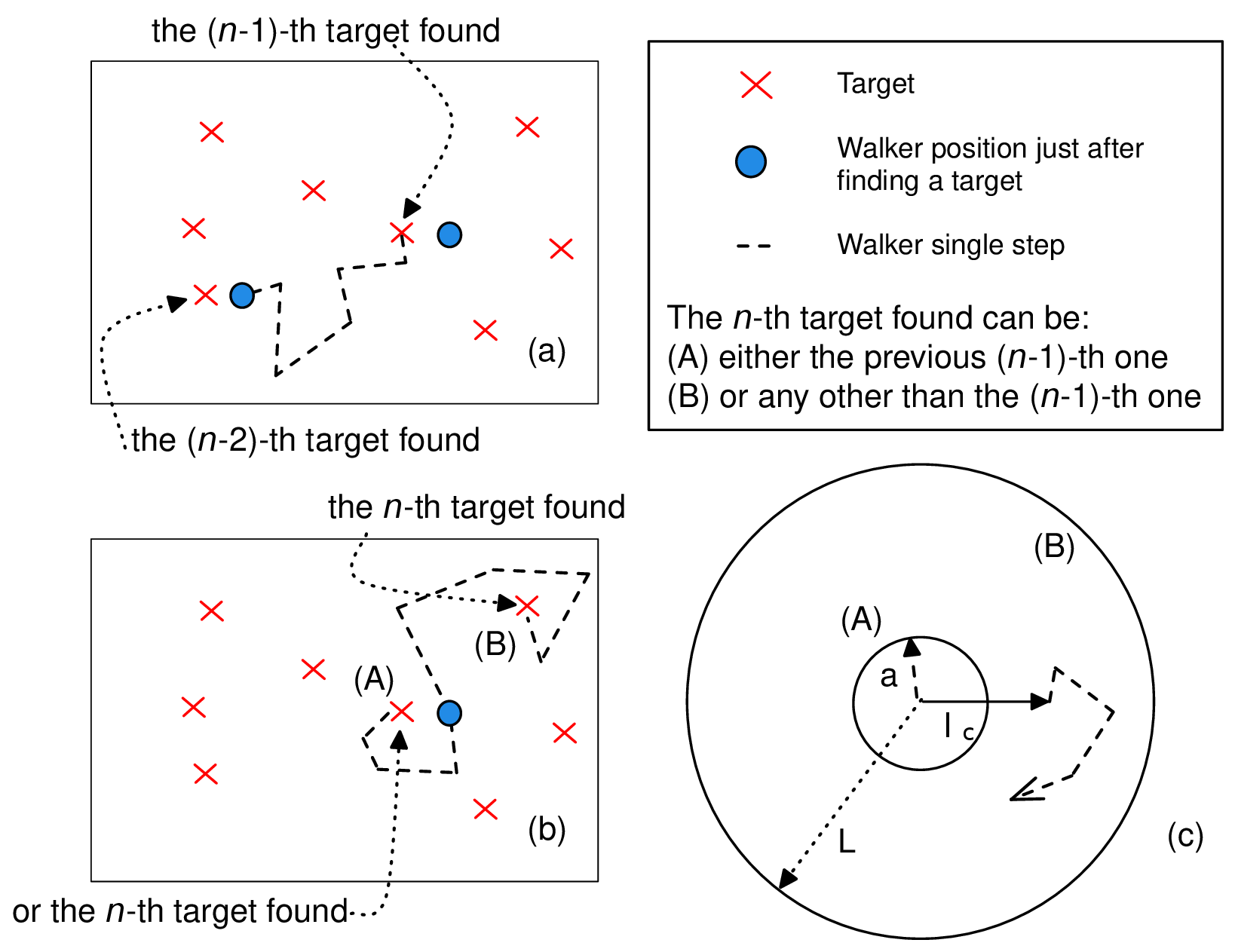}    
    \caption{The random search model is depicted in (a)--(b).
      (a) Always leaving from a position close to the previously
      found target, the walker follows the Lévy walk strategy
      (main text) looking for the next target.
      (b) Once the $n-1$--th target has been found, the $n$--th
      one can be either a revisit to the previous $n-1$--th
      target --- case ($A$) --- or the finding of any other
      than it --- case ($B$).
      (c) A random walker (of similar locomotion rules of
      the random search model) inside an annulus geometry.
      The circumferences ($A$) and ($B$) represent the
      inner and outer annuli of radii $a$ and $L$.
      $l_{c}$ marks the restart point each time the walker
      reaches one of the annuli borders.
      There is a proxy between ($A$) and ($B$) in (b)
      and ($A$) and ($B$) in (c), where being absorbing means finding a target.
}
\label{fig:annuli-schema}
\end{figure}

\begin{figure}[t]
\includegraphics[scale=0.075,trim={0 0 0 20cm},clip]{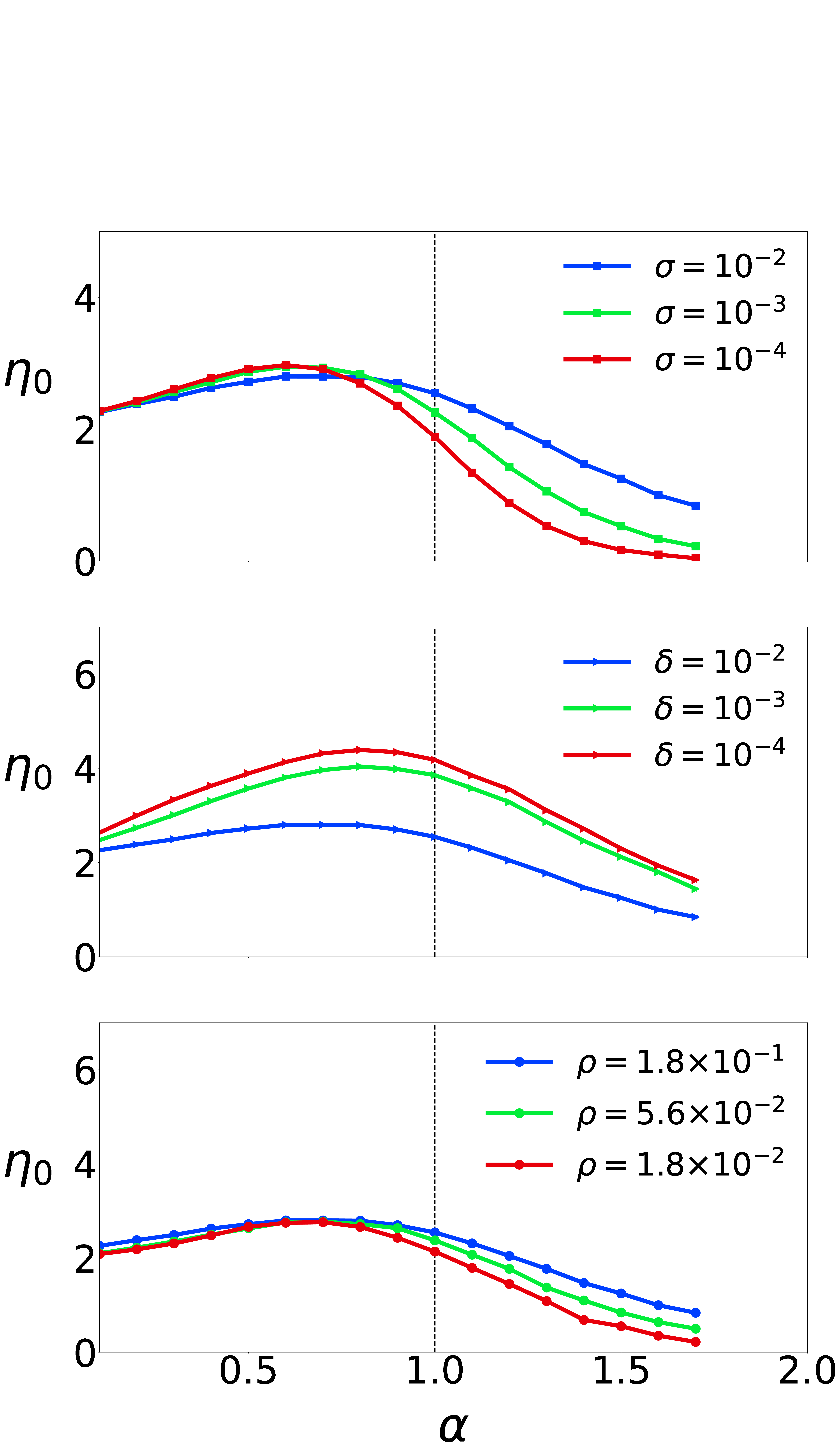}
\caption{Efficiency $\eta_0$ as a function of the Lévy
  index $\alpha$ for different values of
  (a) Lévy scale parameter $\sigma$ ($\rho=5.6\times10^{-2}$
  and $\delta=10^{-2}$),
  (b) the relative distance from the inner radius
  $\delta$ ($\sigma=10^{-2}$ and $\rho=5.6\times10^{-2}$),
  and
  (c) the effective density $\rho$ ($\sigma, \delta=10^{-2}$).
  The vertical line $\alpha=1$ is just guide for the eye.}
\label{fig:eff-delta-rho-sigma}
\end{figure}

In 1D, the foraging model described above can be solved
analytically for the target density going to
zero~\cite{nature1999,pre-sergey,buldyrev2001}.
Specifically, the ``mean-field'' treatment in
Ref.~\cite{nature1999} was later rigorously established
in Refs. \cite{pre-sergey, buldyrev2001} using fractional
differential equations with a Riesz kernel.
From this  framework, one finds that inverse square Lévy
walk searches are optimal in the limit of very low target
density (ideally vanishing).

In 2D, the problem is substantially more difficult.
Numerical simulations have strongly suggested that the 1D
result for the optimality of inverse square Lévy walks
extends to 2D and 3D (see, e.g.,
Ref.~\cite{foraging-book, pnas2008, fpgm2008} and references
therein).
At present, the rigorous analytical treatment of the general
2D problem is considered to be extremely hard, with the
eventual exception of the limit that corresponds to the
foraging with revisits (which just happens to be the most
important case in real applications) as 
discussed in~\cite{comment}.

{\em Search inside annuli and spherical shells}:
A natural way to tame the original problem, while still
retaining most of its important ingredients, is to consider
the simpler situation of a Lévy walker inside a 2D annulus,
with absorbing boundaries at the inner (border $A$) and
outer (border $B$) radii,
(c) in Fig. \ref{fig:annuli-schema}.
Note that to find either the previously found target
(the closest one) or any other farther away bears a close
relation to reaching either the inner circle border
or the outer circle border, Fig.~\ref{fig:annuli-schema}.
Indeed, the inner circle or sphere represents the
previously visited target, whereas the outer circle
or sphere represents all the other targets averaged out.
{Specifically, in passing from the original foraging problem to the annulus problem,
the distribution of the distances to the nearest target is replaced by
its mean value. Since the distribitution has rotational symmetry, the
mean distance is independent of the direction.  In 2D the locus of
points with fixed distance from the previous target is a circle. In
$d$ dimensions it is a $d-1$-dimensional surface of a $d$-dimensional
sphere. }

The dynamics of this simplified model follows the rules:
\\
(1) The walker always starts (or restarts) at
$(\ell_c, \theta=0)$, close to the border of the inner
circle ($\ell_c - a$ is positive but small compared to
the outer radius $L$).
It then moves between the inner and outer spherical
shells, being absorbed upon hitting any of the two.
\\
(2) At the $j-$th step, the walker follows the
rule (ii) of the random searcher, with the
difference that instead of looking for a target the
step terminates if $\ell_{j}$ is enough to reach one
of the borders, when then (1) resumes.

The FPT efficiency of the above model is characterized
as in Eq. (\ref{eff_ratio}), but with $N$ now being
the number of times the walker is absorbed by the
annuli borders.
Very importantly, it is simple to realize that
there is a direct correspondence between optimizing
the foraging $\eta$ and minimizing the FPT for the
absorbing shells problem. 

\section{Analytical Results} \label{sec-ana}

The asymptotic behavior of our random walk inside annuli or
$D$-dimensional spherical shells of inner radius $a$ can be
obtained analytically in the triple limit
{
\begin{equation}
  l_c\to a, \qquad L  \to \infty, \qquad {s } \to 0~,~~~ s<l_c-a~.
  \label{limites}
\end{equation}
}
where $s$ is the scaling factor of the
distribution of step lengths. As discussed in \cite{comment}, this would
correspond to the crucial limit of foraging with revisits.
{
The limit $\delta \to 0$
means, for the foraging model, that the previously visited target
becomes revisitable again almost immediately. In the annulus model,
this limit correponds to being extremely close to the inner circle
(remembering that the inner radius corresponds to the radius of
detection for the foraging model).  The limit $\sigma\to 0 $ in both
models means that the smallest individual random walk steps goes to
zero. If this limit is not taken, then when for $\delta$ sufficiently
small, the very first Lévy walk step might be larger than the distance
to the previously found target in the foraging model, or else to the
inner radius in the annulus model. Specifically, we need $l_c -a \gg
s$ to avoid the problem with the first step being dominant, and this
condition is equivalent to $\delta\gg \sigma$. 
Finally, the limit $\rho \to 0 $ is
simply reducing the target density to zero in the foraging model and
reducing the equivalent quantity to zero in the annulus model. We have
been careful to take the limits in the right order, i.e., first
$\sigma$, then $\delta$, and finally $\rho$.}

There are many power law tailed distributions, all
of which should lead to similar results
in this triple limit due to the generalized central
limit theorem for Lévy $\alpha$-stable
distributions~\cite{Levy,Kolmogorov}.
The $\alpha$-stable Lévy distribution has the
probability density function
$f(x;\alpha,\beta,d,s)$ given by
\begin{equation}
  f(x;\alpha,\beta,d,s)=
  \frac{1}{2\pi} \int_{-\infty}^{\infty}  \exp[\phi(t)]
\, \exp[- i x t] ~dt,
\label{eq1}
\end{equation}
with
$$\phi(t)=\left \{
\begin{array}{ll}
i t d -
|s \, t|^\alpha \,
\big(1 - i \, \beta \, \sign[t] \tan[\frac{\pi}{2}\alpha]\big),
&  \mbox{for} \ \ \ \alpha \neq 1  \\
i t d - |s \, t|
\big(1 + i \, \beta \, \frac{2}{\pi} \, \sign[t] \,
\ln[|t|]\big),
& \mbox{for} \ \ \  \alpha=1~
\end{array}
\right.
$$
with $d$, $s$ reals and $\beta \in [-1,1]$.
The Lévy index $\alpha\in (0, 2]$ governs the
asymptotic behavior of $f(x;\alpha,\beta,\mu,s)$ in
the form of the power-law tail $\sim 1/x^{\mu}$,
with $\mu=\alpha+1$ for $\alpha<2$.
For $\alpha=2$ one recovers the Gaussian, since then
the second moment is finite and the usual central
limit theorem holds.
Further, $\beta$, $d$ and $s$ represent, respectively,
the distribution asymmetry or skewness, shift or
location and the scaling for the $x$ variable.

In what follows, we assume without loss of generality
that $p(\ell)$ is given by $f(\ell,\alpha,0,0,s)$.
Taking $\beta=d=0$ is justified because the model
should have rotational symmetry (e.g., with $p(\ell)
= p(-\ell)$) and then $s$ can be interpreted as
a width.
With this choice, $\mu = \alpha+1$ for $\alpha<2$.

{ 
We also note that there is more than one way of generating
Lévy walks in higher dimensions.
The method we use here bears resemblance to the uniform model
of Lévy walks considered by Zaburdaev {\it et al.}~\cite{z-prl}.
Further  detailed discussions lie beyond the scope of the present
contribution.
Hopefully, this rather technical aspect will be addressed in a
future study.

In the foraging problem, there is a well-behaved relationship
between the mean free path $\lambda$ between targets and the
target density $\rho$.
With $a$ being the radius of detection, in 2D we have
$2 a \lambda \rho =1$.
In $D$ dimensions, we similarly have
$\rho \sim 1/(a^{D-1}\lambda)$.
Noting that for annuli and shells the mean free path goes
with $L$, we can define the effective density according to
  \be
  \rho =  \frac 1 {a^{D-1} L }.
  \label{eq:density}
  \ee
Equation (\ref{eq:density}) agrees with the definition
for the foraging problem up to a constant factor. 

Based on Fig. \ref{fig:annuli-schema} (c) and Eqs.
(\ref{levy_dist}) and (\ref{eff_ratio}), we can 
then further define --- moreover observing that in
fact $\eta = \eta(\alpha,\delta,\rho,\sigma)$ ---  
that
\begin{equation}
  \delta = \frac{l_c}{a} - 1, \ \ \
  \sigma ={s \over a}, \ \ \
  \eta_0 ={\eta  \over \rho \, a^{D-1}}=\eta L.
  \label{eq:parameters}
\end{equation}

Note that $\eta_0$ is adimensional.
The important limit \mbox{$\delta \to 0$} (in which the
walker is extremely close to the inner annulus of radius $a$)
corresponds to the limit of foraging with revisits.
Therefore, the curvature of the inner circumference can
be neglected:
the walker sees the surface of the inner circle or sphere
as a ``flat wall'' regardless the dimension $D$.
This is the reason why the walker behavior is well
approximated by the one-dimensional theory.
In this way, the rigorous theory of the Riesz
operator~\cite{buldyrev2001} on an interval of length $L$ with
absorbing ends becomes applicable~\cite{comment}.

}

Thence, for $\sigma >\delta$ the efficiency increases when
$\sigma$ decreases because there are fewer large jumps
away from the previous target that makes re-encountering
it difficult \cite{comment}.
When $\sigma\approx \delta$ the efficiency saturates and
should reach its maximum.
In fact, for \mbox{$\sigma\approx\delta\to 0$} we must
have the same scaling behavior as in {$d=1$}.

The $1D$ behavior has been known exactly for a few
decades (for details see Refs.
\cite{pre-sergey,buldyrev2001}).
Extending it to the present case, we get
\begin{equation}
  \eta_0\sim \left\{\begin{array}{ll} \delta^{-\alpha/2},
  &  \alpha <1,
\\
\delta^{-1+\alpha/2}, &  
\alpha>1. \end{array}
\right.
\label{eq-eijrnbienrtgu848}
\end {equation}
Here it is worth presenting a simplified summary of this optimization
result.  On the one hand, for $\alpha>1$ the mean step size is finite
and as $\alpha$ increases, the searcher spends larger and larger
amounts of time backtracking, which increases the time to reach both
outer and inner boundaries.  The minimum time to reach the boundary is
thus given by $\alpha \leq 1$.  On the other hand, for $\alpha<1$ the
mean step size diverges and as $\alpha$ decreases, the probability of
reaching the outer boundary on the first step increases, which reduces
the probability of reaching the inner boundary which is extremely
close.  Hence, the optimal efficiency is given by $\alpha\geq 1$.  The
two inequalities, $\alpha\geq 1$ and $\alpha\leq 1$, are of course
satisfied only if $\alpha=1$.
{Eq.~(\ref{eq-eijrnbienrtgu848}) is the main result
  reported here for the annulus problem.  See appendix B and C for
  very similar results obtained via alternative arguments, for the
  foraging problem and the annulus problem respectively.}

\begin{figure}[t]
\includegraphics[scale=0.079]{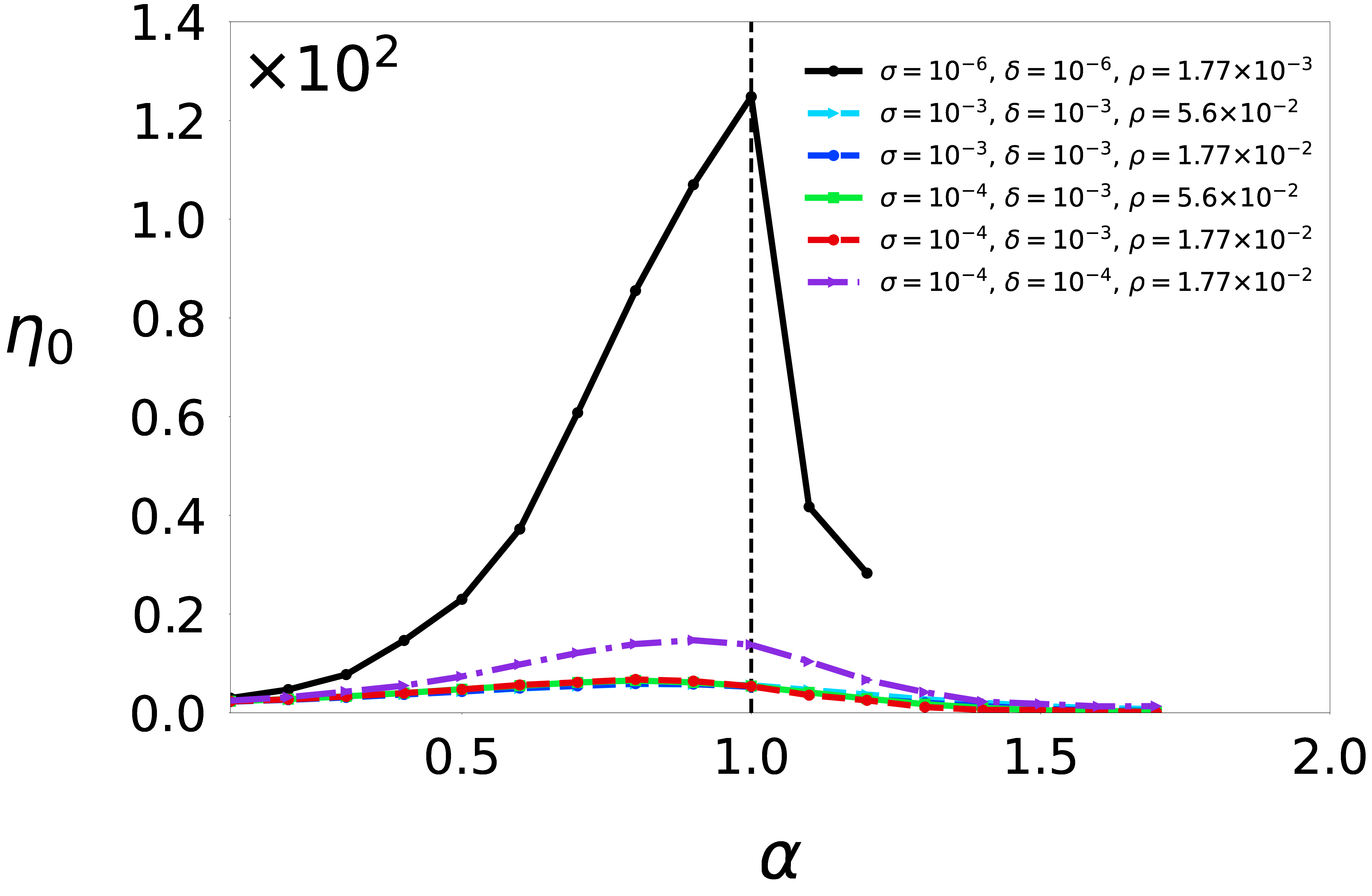}
\caption{Efficiency $\eta_0$ as a function of the Lévy index for
  various values of $\sigma$, $\delta$, and $\rho$ for search inside a 2D annulus.  As these
  parameters tend to zero, the optimal Lévy index goes to 1,
  corresponding to inverse square Lévy walks.  The efficiency $\eta_0$
  is so large for the case $\sigma=\delta=10^{-6}$ and
  $\rho=1.77\times10^{-3}$ (black curve) that the other curves appear
  to be zero on a linear scale.
  For large $\alpha > 1.2$ some points
  are not shown because the computational runtime can become
  extremely large.
  What is important to note, however, is the behavior near the
  peak.}
\label{fig:eff-alpha-1}
\end{figure}

Thus, $\eta_0$ has an arbitrarily strong maximum
at $\alpha=1$ when $\sigma\approx\delta\to 0$ in any
dimension.
This is the key advance reported here.

\section{Numerical Results} \label{sec-num}

In what follows, we verify this theoretical prediction
via numerical simulations. 
Nevertheless, we comment that although a more rigorous
and full mathematical treatment of the $D$-dimensional
is beyond the reach at the present time, still one can
derive approximate analytic solutions.
We refer the interested reader to the Supplementary
Material accompanying the present work (in particular,
see Appendices \ref{s1} and \ref{s2}).

{\em Numerical checks}:
Next we present some numerical simulations to corroborate
our analytical derivations.
First, we consider $\eta_0$ for small values of pertinent
parameters, but not yet fully corresponding to the
limit in Eq. (\ref{limites}).
In Fig. \ref{fig:eff-delta-rho-sigma} we plot $\eta_0$ as
a function of $\alpha$ assuming different
$\sigma$, $\rho$ and $\delta$.
For instance, when $\delta$ and $\rho$ are fixed
and $\sigma$ is decreased, the efficiency $\eta_0$ for
large $\alpha$ decreased, as seen in Fig.
\ref{fig:eff-delta-rho-sigma} (a).
This behavior should be expected since smaller $\sigma$'s
implies smaller step sizes, hence making it less probable
to reach the target in the first few steps.
On the other hand, for $\sigma$ and $\delta$ fixed,
as $\rho$ increases the relative prominence of the peak
for $\eta_0$ increases, Fig. \ref{fig:eff-delta-rho-sigma}
(c), again an expected result.
Actually, an ever-increasing maximum for $\eta_0$ is only 
possible in the limit $\rho \to 0$.
Lastly, by decreasing $\delta$, while keeping $\rho$ and
$\sigma$, we observe an overall increase of the curve
$\eta_0$ versus $\alpha$. 
This agrees with the fact that for smaller $\delta$ the
searcher will more frequently find the closer target
regardless of $\alpha$, thus increasing the total
efficiency.

\begin{figure}[t]
\includegraphics[width=\linewidth]{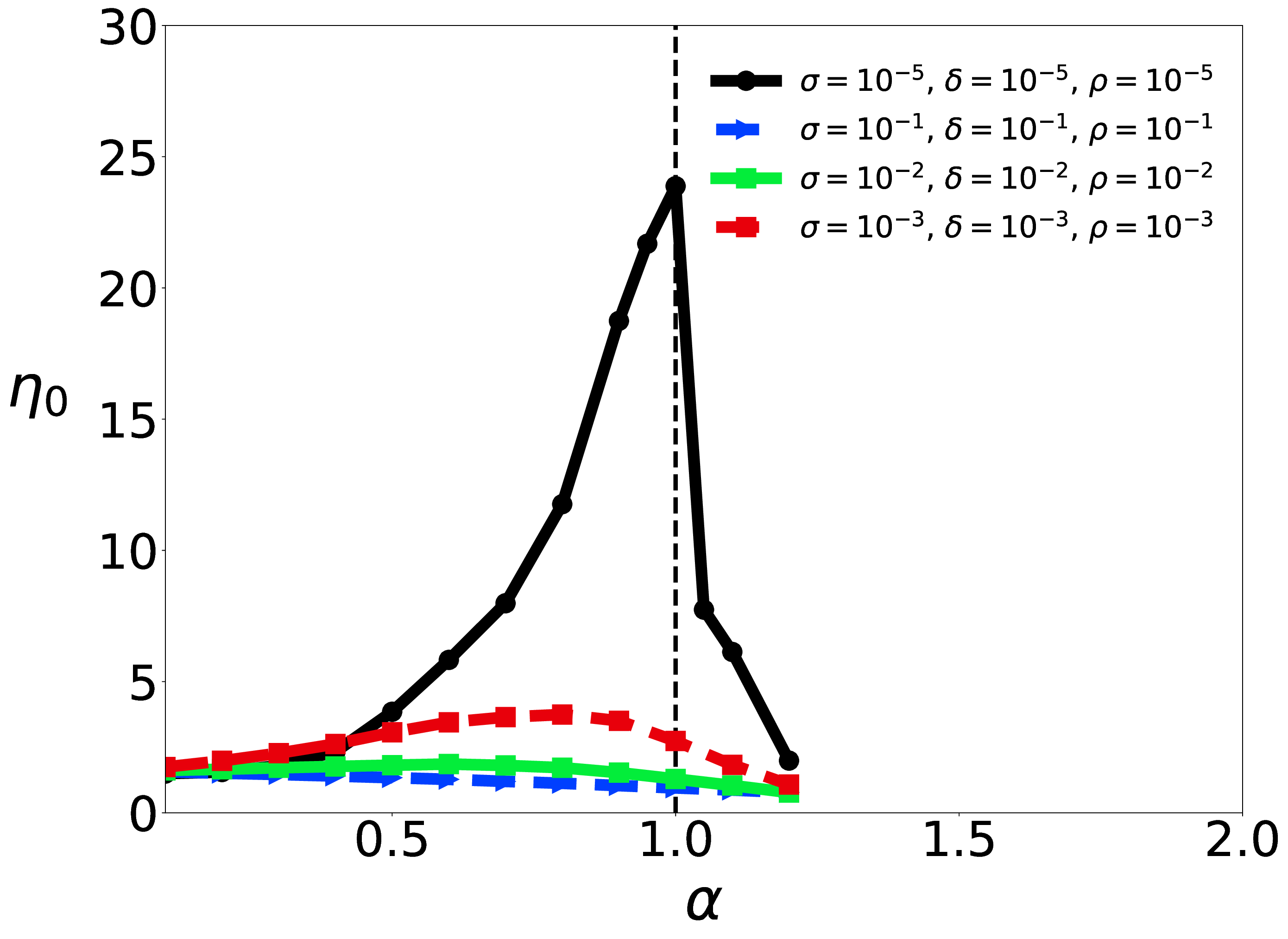}
\caption{Efficiency $\eta_0$ as a function of the Lévy
  index for various values of $\sigma$, $\delta$ and
  $\rho$ for search inside a 3D spherical shell.
  As predicted theoretically, a maximum emerges near
  $\alpha=1$ as $\sigma, \delta, \rho\to 0$.
  Again, for $\alpha > 1.2$ the points are not shown
  because the computational runtime, even more critical
  in 3D.
   }
\label{fig3d}
\end{figure}

If $\sigma$, $\delta$ and $\rho$, Eq. (\ref{eq:parameters}),
go to zero in a proper manner (see Eq. (\ref{limites}))
we should expect $\eta_0$ to become arbitrarily large near
$\alpha=1$.
Exactly as predicted by the theory, Fig. \ref{fig:eff-alpha-1}
shows how the limit $\delta, \sigma, \rho \rightarrow 0$
influences the shape of the efficiency curve.
Observe that the peak shifts to $\alpha=1$ for the case of
search with unrestricted revisits.
Finally,  Fig.~\ref{fig3d} shows that likewise the maximum at
$\alpha=1$ also occurs for search in 3D spherical shells,
in precise agreement with our findings.

\section{Conclusion} \label{sec-con}

We have presented solid analytical results, checked through
numerical simulations, showing that the inverse square Lévy walks
optimize the time to reach the absorbing boundaries of annuli
and spherical shells.
Given the relationship of this problem to the original foraging
model, these results should be expected to extend to the latter
by proxy.
Most importantly, our analysis unveils the real reason for the
optimality of inverse square Lévy walk search strategies for
foraging with unrestricted revisits (hence also non-destructive
foraging) in any dimension.
Regardless of the dimension $D$, the general process essentially
reduces to the well-understood 1D model in the case of scarce
distribution of targets, not unlike how the approximately spherical
earth appears locally flat for small enough organisms.
We hope that such findings can finally settle positively
the key fact in the theory of Lévy random search.

\bigskip

\begin{acknowledgments}
  We thank Sergey V. Buldyrev for very carefully reading our
  manuscript and giving us very helpful feedback. 
  We thank CAPES, CNPq, and FACEPE for funding.
  F.B.  acknowledges support of Grant CGL2016-78156-C2-1-R
  from MINECO, Spain.
\end{acknowledgments}

\appendix

\section{The foraging problem revisited}
\label{s1}
The foraging problem in 1D is equivalent to a Lévy walker inside an
interval with absorbing boundaries.
As explained in the main text, the one-dimensional case was
rigorously solved in 2001, see Refs. [15,16] (hereafter all cited
references refer to those in the main text).
In higher dimensions, however, the mathematical difficulties seem to be
extremely high, probably demanding the development of new methods
and tools for a proper solution.
As far as we know, to date the only exception is the triple
limit discussed in the main article.
However, it is possible to treat the problem approximately, thus
obtaining some insight into the main physically important features
determining the optimal search strategy.
So, in the following we give such an approximate treatment.

\begin{figure}[t!]
\includegraphics[width=0.5\textwidth]{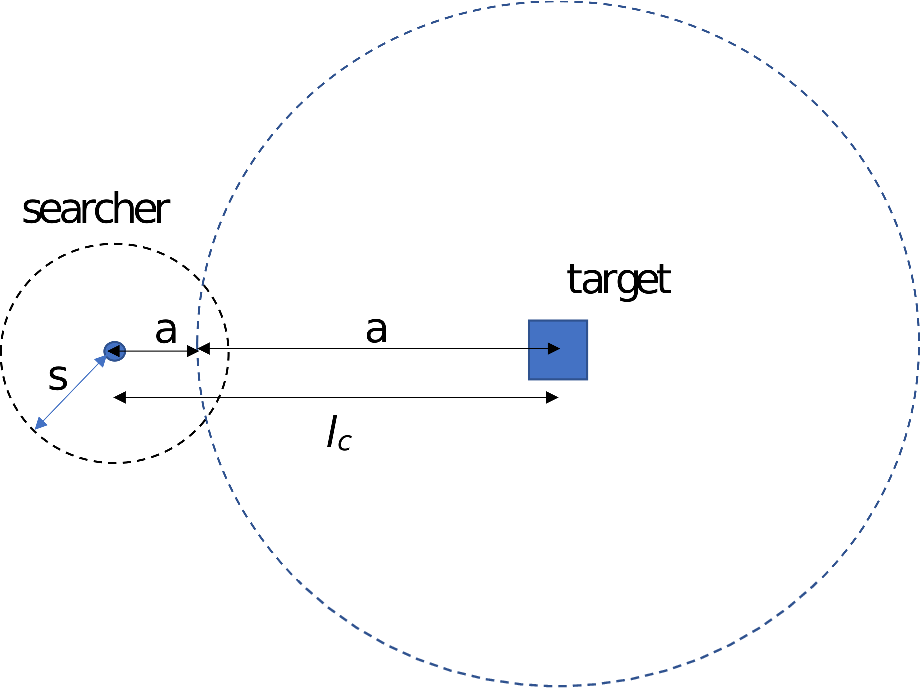}
\caption{The spatial disposition and relevant distances
  of the searcher (dot) and nearest target (blue square).
  The searcher's detection radius is $a$, the initial distance
  from the searcher to the target is $l_c$, and $s$ is the
  scaling factor of the distribution of step lengths.}
  \label{spatial-disposition}
\end{figure}

In Fig. \ref{spatial-disposition} we show the initial condition of a non-destructive
search in 2D.
We define the dimensionless parameter $\delta = (l_c - a)/a$,
with $l_c$ the searcher initial distance to the nearest target
and $\delta a$ its distance to the circumference of radius~$a$
around the target.
We assume the typical length scale $(\rho \, a)^{-1}$ calculated
for a L\'evy walker in 2D, Ref.~\cite{ref1},
so that $\eta = \eta_0 \, \rho \, a$.
Comparing with Eq.~(5) of Ref.~\cite{ref1},
we have that $\eta_0 = K_d/a$ (the notation $K_d$ is
borrowed from Ref.~\cite{ref1}, it represents
the efficiency gain).
The dimensionless searching efficiency is
${\eta_0 (\alpha, \delta, a, s) = f / \langle L \rangle}$,
where $f(\alpha, a, s)$ has dimensions of length
(leaving~$\eta_0$ dimensionless)
and $\langle L \rangle (\alpha, \delta, a, s)$ is the
average distance traversed until the the encounter of a target.  
We shall calculate $\eta_0$ when the search is nondestructive,
$\delta \to 0$, and in the scarce regime, $\rho \to 0$.

For $\delta$ and $\rho$ going to zero, in 2D the encounter of
the very close target (hereafter CT) essentially
determines~$\eta_0$. 
In fact, in the 2D scarce regime, the probability for the
faraway target (FT) --- at a typical distance $\lambda
\gg \delta a$ --- to be the first one to be found is much
lower than $[\delta a/\lambda]^{\alpha/2}$, the corresponding
probability in 1D. 
Thus, in 2D with $\delta \to 0$ and $\rho \to 0$, to first
order one should be concerned essentially with the finding of
the CT. 
Now, to reach this nearby target in the first walk step we
need ${\sigma = s/a \approx \delta \to 0}$, with $s$
the scale of the $\alpha$-stable Lévy distribution (see
main text).
However, if the searcher eventually misses the CT
in the very first step, the next few successive steps
still will lead to the CT provided the searcher does
not move away too much from the CT location, say by
keeping wandering around within a small region of radius
$n \, a$ for $n$ few units, see Fig. \ref{spatial-disposition}. 

Therefore, it is a good approximation to assume that in
the above mentioned small region, the fractional diffusion
equation that governs the L\'evy searcher dynamics leads
to solutions displaying basically the same qualitative
behavior in 1D and 2D.
This way, for $\delta \to 0$,  $\rho \to 0$ in 2D
we have
$\eta_0 \sim f / (\langle n \rangle_{\mbox{\tiny 1D}} \,
\langle \ell \rangle )$,
where the 1D result for a Lévy walker starting from a
distance $\delta \, a \to 0$ to the absorbing CT
within a distance $n \, a$ can be
approximated as in Ref.~\cite{nature1999}, or
\begin{equation}
  \langle n \rangle_{\mbox{\tiny 1D}}
  \sim \left ( \frac{ \delta \, n \, a^2 }{ s^2 } \right )^{\alpha / 2},
\nonumber
\end{equation}
and 
\begin{equation}
\langle \ell \rangle \sim s \left [ \left ( \frac{ \delta \, a }{ s }
 \right )^{1-\alpha} + b \right ],
\nonumber
\end{equation}
where $b \sim 1$.
We thus get
\begin{equation}
  \frac{\langle n \rangle_{\mbox{\tiny 1D}} \,
    \langle \ell \rangle}{f} \sim \frac{a}{f} \,
  \delta^{1-\alpha/2},
\ \ \mbox{with} \ \ \delta \to 0,~~\rho \to 0,~~\alpha > 1
\nonumber
\end{equation}
and 
\begin{equation}
  \frac{\langle n \rangle_{\mbox{\tiny 1D}} \,
    \langle \ell \rangle}{f}
  \sim
  \frac{a}{f} \
  \left ( \frac{a}{s} \right )^{\alpha - 1}
  \, \delta^{\alpha/2},
\ \ \mbox{with} \ \ \delta \to 0,~~\rho \to 0,~~\alpha < 1.
\nonumber
\end{equation}

Then, we obtain in the non-destructive $(\delta \to 0)$
scarce ${(\rho \to 0)}$ regime that~$\eta_0$ scales
with~$\delta$ in the form
\[ \eta_0 \sim  
  \begin{cases}
   \, \delta^{-1+\alpha/2} & , ~~\alpha > 1,\\
   \, \delta^{-\alpha/2} & , ~~\alpha < 1.
  \end{cases}
\]

We once more shall stress that this is an approximate
calculation.
A far more grounded procedure is developed in Refs.~[15,16].
Nonetheless, the above analysis gives the same result as the
more rigorous approach in Ref.~\cite{comment}.

\makeatletter 
\makeatother

\section{An approximate analytical treatment for the
  annulus problem} \label{s2}

The 1D equivalent of an annulus or spherical shell is of
course just an interval, which has already been discussed
in the main text.
Moreover, the foraging problem in 1D is identical to the
problem of the walker inside an interval (with absorbing
boundaries).
But although in higher dimensions these two models are not
exactly mapped into each other, they are still very
closely related.

In the same spirit of the foraging problem in Appendix \ref{s1}, below we
present an approximate solution for the absorbing annulus
model.
It is especially noteworthy that this treatment gives the
same answer for the optimality of inverse square Lévy
walks in the triple limit $\sigma, \rho, \delta~\to 0$
obtained in the main text through mathematically more
well founded arguments.
Nonetheless, an interesting aspect of the approach
below is that it is a kind of mean field method for
the present problem.

The searcher starts from a distance $l_{c}$ of the center.
For $\delta=\left(l_{c}-a\right)/a$ as before, we
are interested in the $l_{c}\rightarrow a\ll b$ limit, i.e.,
$\delta\rightarrow0^{+}$ for which the probability of
the walker reaching the border $r=b$ is very small,
$P \ll P_{\text{1D}}$.
Note that we can approximate
$P_{\text{1D}} = \left[\left(l_{c}-a\right)/
  \left(b-l_{c}\right)\right]^{\alpha/2}$
as the probability of finding the
distant site in 1D, with $P\left(l\right)$ power-law
with $\mu=\alpha+1$ or with a Lévy with index $\alpha$.
The efficiency can then be written as
{
\begin{equation}
  \eta \approx {1 \over \langle n \rangle \langle | l | \rangle }
  ~.
\end{equation}}

Thus, we focus in the case $\delta\rightarrow0^{+}$
for the encounter with the $r=a$ ring and take into
account only the walks that do not stray too far
from the $r=a$ ring.
In other words, we restrict the random walk to the region
$r < \gamma \, a$, where $\gamma$ must be of the order of
unit.
In such framework, the average number of steps is
fairly given by the 1D result, but with
$\lambda \rightarrow \gamma \, a$.
Using
$\eta\backsimeq1/
\left[\left\langle \eta\right\rangle \left\langle
  \left|l\right|\right\rangle \right]$,
we obtain
\[
\eta\sim\begin{cases}
\delta^{-\alpha/2}, & \alpha<1,\\
\delta^{\alpha/2-1}, & \alpha>1.
\end{cases}
\]
which coincides with the 2D result in Ref.~\cite{comment}
for the $\delta\rightarrow0^{+}$ limit
(assuming a Poissonian distribution of targets
whose density $\rho\rightarrow0$).
That is
\begin{align}
  \left\langle \eta\right\rangle _{\text{1D}} & =
  f(\alpha)\left[\frac{x_{0}
      \left(\lambda-x_{0}\right)}{s^{2}}\right]^{\alpha/2}
  \rightarrow
  \nonumber \\
  \left\langle \eta\right\rangle &
  \sim f_{\text{2D}}(\alpha)
  \left[\frac{\delta a\left(\gamma a-\delta a\right)}{s^{2}}
    \right]^{\alpha/2},
\end{align}
    {
where $s$ is the scale parameter of the Lévy distribution $P(\ell)$,
defined from its characteristic function $\bar{P}(k) =
e^{-|sk|^\alpha}$ (see also $\phi(t)$ in the main text).}

For a power law instead of a Lévy stable distribution, 
we can take $s=l_{0}$.
The function $f(\alpha)$ in 1D depends on whether
$P\left(l\right)$ is Lévy or power-law, e.g.,
$f(\alpha) = 1/\Gamma\left(\alpha+1\right)$
in the Lévy case (for $\Gamma$ the Gamma function).
But for purposes of scaling laws of
$\left\langle \eta\right\rangle $
when $\delta\rightarrow0^{+}$, this pre-factor is not
important once it does not depend on $\delta$.

Lastly, to compute $\left\langle \left|l\right|\right\rangle $
we also can adapt the calculations for 1D by
just supposing walks in the region $r < \gamma \, a$.
We can approximate following Ref.~\cite{nature1999}
\begin{equation}
  \left\langle \left|l\right|\right\rangle
  \sim c\left[\left(\frac{\delta \, a}{c}\right)^{1-\alpha}
    g\left(\eta,\alpha\right)+h\left(\alpha\right)\right].
\end{equation}
Here the functions $g$ and $h$ do not depend on $\delta$.
Thence
\begin{equation}
  \left\langle \eta\right\rangle \left\langle
  \left|l\right|\right\rangle \sim f_{\text{2D}}
  \left[\frac{\delta \, a^{2}
  \left(\gamma-\delta\right)}{c^{2}}\right]^{\alpha/2} c \
  \left[\left(\frac{\delta a}{c}\right)^{1-\alpha}g+h\right].
\end{equation}
 Taking the limit $\delta\rightarrow0^{+}$
\[
\left\langle \eta\right\rangle
\left\langle \left|l\right|\right\rangle \sim\begin{cases}
\delta^{\alpha/2}, & \alpha<1,\\
\delta^{-\alpha/2+1}, & \alpha>1,
\end{cases}
\]
by retaining only the dependence of
$\left\langle \eta\right\rangle
\left\langle \left|l\right|\right\rangle $ on $\delta$.

Remarkably, such straightforward and easy-to-understand 
considerations yields a similar result to the
procedure followed in the main text.  
{Recall that in the
    main text, the steps were as follows: 1. The 2D-search-inside-the
    annulus problem is first setup. This general problem has not been
    solved. 2. The limit it taken of $\sigma\to 0,~ \delta to 0,
    ~\rho\to 0$. In this limit, the problem becomes effectively one
    dimensional. 3.  Eq. (\ref{eq-eijrnbienrtgu848}) is thus obtained
    in this limit.}

But of course, from a fundamental point of view the
findings in the main text represent a much more
important and rigorous achievement.

\section{Analytical solution for $\alpha\rightarrow0$} \label{s3}
In the limit where $\alpha\rightarrow 0$, the searcher reaches either
the outer ring or the inner ring in typically just a single step of
size$\ell$. We will try to map the step size $\ell$ into the
parameters of concentric rings. For this, we will use the law of 
cosines:
$L^{2}=l_{c}^{2}+\ell^{2}-2l_{c}\ell\cos\left(\pi-\theta\right)$, with
$\theta$ being the step direction.

If {$L \gg l_{c}$}, we have that $\ell=l_{c}\cos\left(\pi-\theta\right)+\sqrt{l_{c}^{2}\cos^{2}\left(\pi-\theta\right)+L^{2}-l_{c}^{2}}$
, for $0\leq\theta<\theta_{\text{max}}$, where 
\begin{equation}
\sin\left(\pi-\theta_{\text{max}}\right)=\frac{a}{l_{c}}\Rightarrow\theta_{\text{max}}=\pi-\arcsin\left(\frac{a}{l_{c}}\right).
\end{equation}

For $\theta_{\text{max}}\leq\theta\leq\pi$, we have that
$a^{2}=l_{c}^{2}+\ell^{2}-2l_{c}\ell\cos\left(\pi-\theta\right).$ If
$a<l_{c}$,
\begin{equation}
\ell(\theta)=l_{c}\cos\left(\pi-\theta\right)-\sqrt{l_{c}^{2}\cos^{2}\left(\pi-\theta\right)+a^{2}-l_{C}^{2}}
\end{equation}
. Let $p\left(\theta\right)$ be the PDF of the turning angle $\theta$,
such that $\int_{0}^{2\pi}p\left(\theta\right)d\theta=1$. This way$\left\langle \ell\right\rangle =\int_{0}^{2\pi}\ell\left(\theta\right)d\theta$.
Substituting in this integral $\ell\left(\theta\right)$ and assuming
$p\left(\theta\right)=1/2\pi$, it follows that
\begin{equation}
\begin{aligned}\left\langle \ell\right\rangle  & =\frac{1}{\pi}\int_{0}^{\theta_{\text{max}}}\sqrt{L^{2}-l_{c}^{2}+l_{c}^{2}\cos^{2}\left(\pi-\theta\right)}\\
 & -\frac{1}{\pi}\int_{\theta_{\text{max}}}^{\pi}\sqrt{a^{2}-l_{c}^{2}+l_{c}^{2}\cos^{2}\left(\pi-\theta\right)}d\theta,
\end{aligned}
\end{equation}
which leads to 
\begin{equation}
\begin{aligned}
\left\langle \ell\right\rangle  & =\frac{1}{\pi}\int_{0}^{\theta_{\text{max}}}\sqrt{1-\frac{l_{c}^{2}}{L^{2}}\sin^{2}\theta}d\theta\\
 & -\frac{a}{\pi}\int_{\theta_{\text{max}}}^{\pi}\sqrt{1-\frac{l_{c}^{2}}{a^{2}}\sin^{2}\theta}d\theta.
\end{aligned}
\end{equation}
 Since a second-order elliptical integral function is given by 
\begin{equation}
E\left(\varphi,m\right)=\int_{0}^{\varphi}\sqrt{1-m\sin^{2}\theta}d\theta,
\end{equation}
 then 
\begin{equation}
\begin{aligned}
\left\langle \ell\right\rangle  & =\frac{L}{\pi}E\left(\theta_{\text{max}},\frac{l_{c}^{2}}{L^{2}}\right)+\frac{a}{\pi}E\left(\theta_{\text{max}},\frac{l_{c}^{2}}{a^{2}}\right)\\
 & -\frac{a}{\pi}E\left(\pi,\frac{l_{c}^{2}}{a^{2}}\right).\label{eq:solution-mean-ell}
\end{aligned}
\end{equation}

Finally, since 
\begin{equation}
\eta=\frac{1}{\left\langle \ell\right\rangle },
\end{equation}
 we obtain the normalized efficiency as 
\begin{equation}
\overline{\eta}=\frac{\eta}{\rho a}=\frac{\eta\pi L^{2}}{a},
\end{equation}
 this efficiency relation, coupled with (\ref{eq:solution-mean-ell}),
 allows comparing the efficiency obtained in this section with that
 obtained in the main text, {namely for $\eta_0$.}

\section{Renormalization group derivation for the efficiency
  scaling with $\rho$} \label{s4}

As it should be clear from the main text, the association
between the foraging and annulus models arises from the fact
that the latter represents one target (CT) at the center,
the origin, and the mean of all other targets (the FTs) are
at the outer radius $b$, so that $\lambda = b$.
Rigorously one should have $\lambda =b-a$,
but we can neglect $a$ in the low-density limit since then
$b \gg a$.

Consider now an annulus with outer radius $L_0=1$ and
inner radius $a$.
In 2D we will define the effective density as usual, according to 
\be
\rho = 1 /(2 \, a \, \lambda),
\ee
where $\lambda$ is the mean free path $\lambda=b$.
Further, let $T$ denote the mean first passage time and
$v=1$ the adimensional unity velocity, so that 
\be
\eta =  \frac 1 T.
\ee

If now we have an absorbing annulus system of outer radius
$L \gg L_0$, obviously its $\rho$ will be much smaller than
the density corresponding to $L_0$.
We then can use renormalization to map the two cases
as the following.
We set $\phi = L/L_0$ and suppose the mapping 
\bea
L &\mapsto& L/\phi = L_0 = 1 \label{eq-L} \\
\lambda &\mapsto& \lambda/\phi = L_0 = 1 \\
a &\mapsto& a/\phi \\
\rho &\mapsto&  \rho \, \phi^d \\
s &\mapsto& s/\phi \\
l_c &\mapsto& l_c/\phi \\
\delta &\mapsto& \delta \\
T &\mapsto& T/\phi \\
\eta &\mapsto& \eta \, \phi  ~. \label{eq-eta}
\eea

Now, assume that the problem for $L_0$ has been solved.
By recalling  that 
\be
\eta_0={\eta L},
\ee
we have
\be
\eta_0 \mapsto \eta_0 \, [ \phi / \phi ] = \eta_0.
\ee
Since
\be
\eta  = \eta_0 \, \rho \, a^{d-1} 
\ee
and given that $\eta_0$ is an invariant of the
renormalization group flow map, we can expect to find 
$\eta \sim \rho$ for fixed $a$ and $\delta$
(so fixed $l_c$).

The above is a non-rigorous scaling argument for the claim 
first published in Ref.~\cite{ref1} (and stated as proposition (i) in 
Ref.~\cite{comment}), namely, that the efficiency $\eta$ is linear
in the density asymptotically. {Note that only Eqs. (\ref{eq-L}) and (\ref{eq-eta})
are needed, the other mappings are shown only for greater clarity.
    }

\section{Algorithm}\label{s5}

\begin{figure*}[t]
\includegraphics[width=20cm, height=15cm]{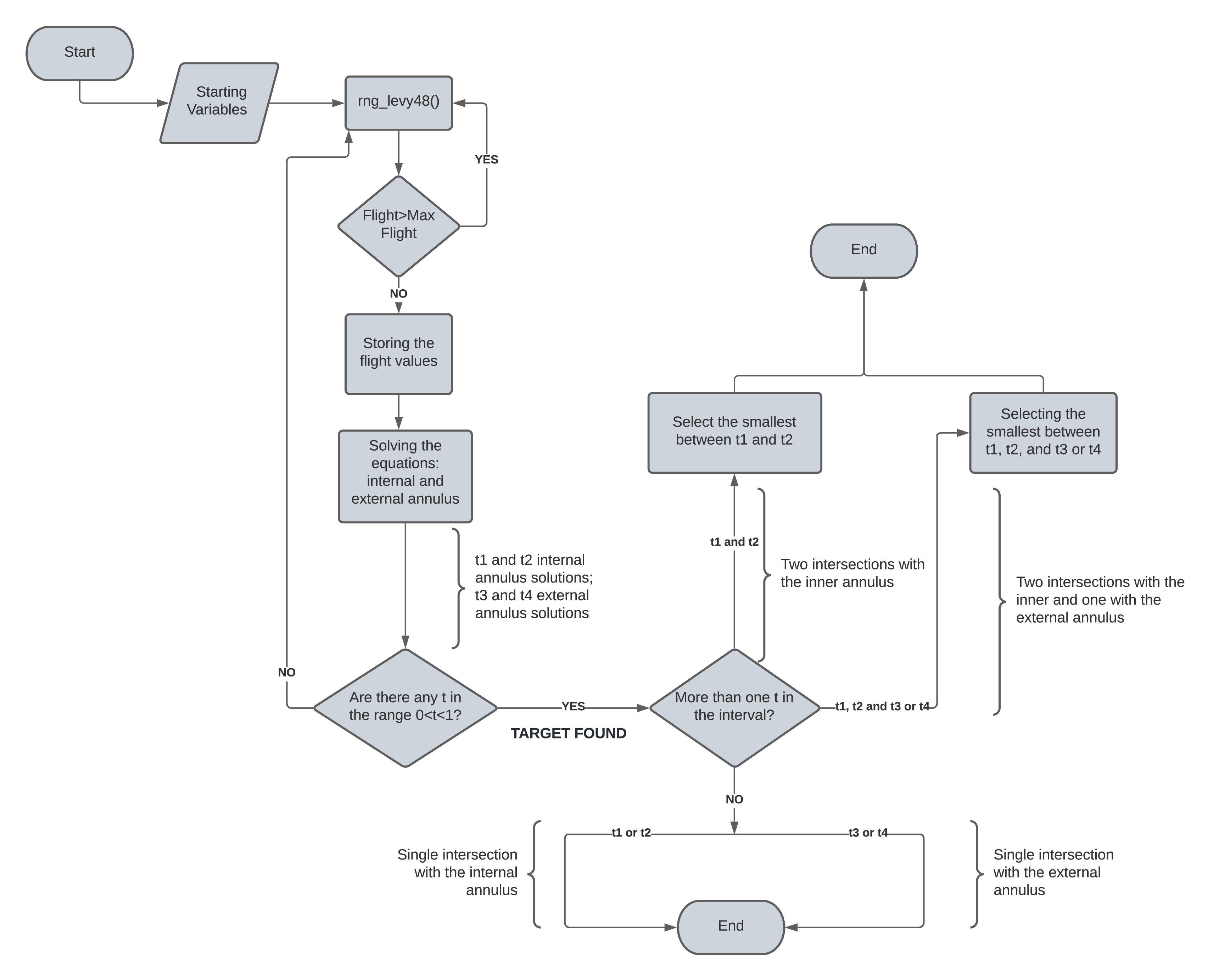}
\caption{Flowchart of the search algorithm for concentric
  annuli with absorbing boundaries.}
\label{fig-flowchart}
\end{figure*}

\begin{shaded}
\samepage{
\begin{verbatim}
double rng_levy48(double alpha, double rr){
    double ee, phi;
    double mu=alpha;
    double mu1=mu-1;
    double xmu=1/mu;
    double xmu1=xmu-1;
    phi=(drand48()-0.5)*PI;
    ee=-log(drand48());
    return rr*sin(mu*phi)/pow(cos(phi),xmu)
             *pow(cos(phi*mu1)/ee,xmu1);
  }
\end{verbatim}}
\end{shaded}

The random variables are generated from the Lévy $\alpha$-stable
distribution with asymmetry parameter $\beta=0$ and zero mean,
also the scale is $s$=\texttt{rr}.
The simulations were performed with a homemade code written in C
(the language C has been chosen due to speed).
The Lévy distributed random numbers were generated using the
C code displayed in the chart.

For the computational simulations, the 2D algorithm checks,
at each walk, whether or not the Lévy searcher has intercepted
the inner or the outer annuli.
Figure \ref{fig-flowchart} depicts the algorithm flowchart.
We use as parameterization $r\left(t\right)=r_{0} + l \, t$,
where $r_{0}$ is the walker starting position.

The algorithm consists of repeating many iterations of the walker
always starting a distance $l_c$ from the center, and performing
successive Lévy walk steps one at a time.
For each step, possible intersection points of the trajectory
and the inner or outer annuli (or shells) are calculated, by
simultaneously solving the equations for the line of the
trajectory and the circles or shells, considered as quadratic
equations for $D$-dimensional conic sections.
The full procedure allows accessing intermediate values between
the points $r_{0}$ and $r_{f} = r \left(t=1\right)$ from the
function $r\left(t\right)$, associating this parameterization
to the equations of each annulus.
We derive the  $t$ value necessary for the intersection from
\begin{align} t_{\text{inner}} &
  =\frac{1}{2A}\left(-B\pm\sqrt{B^{2}-4A
    \left(C-a^{2}\right)}\right),
  \\ t_{\text{outer}}
  &
  =\frac{1}{2A}\left(-B\pm\sqrt{B^{2}-4A
    \left(C-L^{2}\right)}\right),
\end{align}
with $A=2l^{2}$, $B=2\left(x+y\right)l$ and
$C=x^{2}+y^{2}$.
Therefore, we have two solutions for the inner and
two for the outer annulus.
Those with values in the interval $\left(0,1\right]$
indicate that there were one or more intersections.
We choose the lowest $t$ in the interval to compute the
distance traveled $d=lt$.
At the end of the flight, the distance traveled is counted,
and there are three possible subsequent actions:
(i) if there is an intersection, the searcher will return to
the starting point of the simulation,
$\left(x,y\right)=\left(l_{c},0\right)$;
(ii) if there is no intersection, the next flight will
depart from $r_{f}$;
(iii) if the total distance value exceeds a certain threshold,
then the simulation ends.

The 3D version of the code was written by adapting
the 2D code.

\end{document}